\documentclass{appolb}
\usepackage{epsfig}

\begin{document}
\title{Chiral restoration phase transition within the quarkyonic matter%
\thanks{Presented at "Three Days on Quarkyonic Island", 19-21 May, 2011,
 Wroclaw, Poland}%
}
\author{L. Ya. Glozman, V.K. Sazonov and R. F. Wagenbrunn
\address{Institute for Physics, Theoretical Physics Branch,
University of Graz, Universit\"atsplatz 5, A-8010 Graz, Austria}
}

\maketitle
\begin{abstract}
We overview a possible mechanism for confining but
chirally symmetric matter at low temperatures and large densities.
As a new development we employ a diffused quark Fermi surface
and show that such diffusion does not destroy  possible
existence of a confining but chirally symmetric matter at
low temperatures and large density.
\end{abstract}
  
\section{Introduction}

What happens with confinement and chiral symmetry
dynamical breaking in the low temperature matter
at large density, i.e., which phase will take place just
next to the liqud nuclear matter? It was a general belief
in the past that deconfinement and chiral restoration
phase transitions (crossovers) should coincide. A rationale
for such an expectation was the view that the hadron mass 
generation in the confining mode proceeds  through the
chiral symmetry breaking in the vacuum and that the hadron
mass is practically completely due to the quark condensate
of the vacuum. Consequently, beyond the chiral restoration
line hadrons cannot exist and the QCD matter should be in
the plasma form.

In the vacuum we know from the 't Hooft anomaly matching
conditions that indeed in the confining mode chiral symmetry
must be realized in the Nambu-Goldstone mode \cite{anomaly}.
The essence of the argument is that in the vacuum the anomaly
can be saturated only via the massless Goldstone particle
associated with the chiral symmetry dynamical breaking.
However, in the two-flavor baryonic medium the anomaly can 
be trivially saturated with the baryon - baryon hole massless 
excitations and the existence of the massless pion is apriori
not required. 

Another argument was that, according to Casher \cite{Casher},
chiral symmetry breaking is required for quarks to be
confined. Then, naively, hadrons with nonzero mass
cannot exist in a world with unbroken chiral symmetry.
However, the Casher argument is not general and
can be easily bypassed \cite{G1}. Recent lattice simulations
have convincingly demonstrated that in the world without the
low-lying eigenmodes of the Dirac operator (i.e., with the
artificially restored chiral symmetry) hadrons still exist and
confinement persists \cite{Lang}.

In the large $N_c$ world, confinement survives in a cold
matter up to arbitrary large density \cite{pisarski}. In this
case nothing can screen the confining gluonic field and
gluodynamics is the same as in the vacuum. The allowed excitation
modes are of the color singlet nature and it is possible
to define the quarkyonic matter as a dense matter with confinement.
Then a key question is at which density and how could 
a dense cold matter be deconfined.

At $T=0$ deconfinement could happen through the Debye
screening of the confining gluon field: A gluon creates
the quark - quark hole pair that again annihilates into
a gluon. If this vacuum polarization diagram is finite,
then at some density  confining gluons
will be completely screened and the deconfinement would appear.
However, in the confining mode the energy of the colored
quark - quark hole pair is infinite. The allowed excitations
in the confining mode are the color singlet excitations like
the baryon - baryon hole pairs, etc. These excitation modes
cannot screen the colored gluonic confining field. In this
sense, the $T=0$ physics is very different from the high temperature
physics where  deconfinement (screening) proceeds via the incoherent
thermal gluonic loops.

One could expect that the deconfinement in a dense
medium should happen due to per-location of baryons.
Such a reasoning is too naive, however, as the per-location
does not yet imply the screening of the confining gluonic field. 

Lattice simulations suggest that for the $N_c=2$ QCD  at low
temperatures  deconfinement happens at densities of
the order of 50 times the nuclear matter density \cite{Hands}. Since
the $N_c=3$ world is just between two known cases $(N_c =2; N_c=\infty)$,
we expect that at $N_c=3$ the deconfinement at low
temperatures happens at extremely large densities, much
larger than can be achieved in laboratories or in
neutron stars.

By definition the quarkyonic matter is a dense cold matter
with confinement. 
What happens with chiral symmetry breaking
in a very dense cold matter with confinement? Is it
possible to have a chiral symmetry restoration phase
transition in a mode with confinement both below and
above the chiral phase transition?

We cannot solve QCD at large density and the only
way to address this interesting question is to study such a possibility
within a confining and chirally symmetric model. Obviously,
such  model must provide dynamical chiral symmetry breaking
in the vacuum. The simplest possible model that satisfies all
requirements is the model of refs. \cite{Y}. It is assumed
within the model that there is the linear instantaneous confining
potential of the Coulomb type.
Such a potential is indeed observed in Coulomb gauge QCD
 lattice simulations \cite{lattice} or within
the variational approach \cite{var}.
 Then the chiral symmetry breaking
is obtained from the solution of the gap equation. Given the
quark Green function derived from the gap equation, one is able
to solve the Bethe-Salpeter equation for mesons or the corresponding
equations for baryons. An important aspect of this model is that
it explicitly demonstrates the effective restoration of chiral 
symmetry in hadrons with large angular momenta \cite{G2,WG}.
This means that the mass generation mechanism in these hadrons
is not through the chiral condensate of the vacuum. Then, it
is clear, that there is a chance to obtain within this setup
a chirally symmetric but confining matter.

The latter question was addressed in refs. \cite{GW}.
Assuming a liquid phase, i.e., that the translational and
rotational symmetries are intact (as it is in the nuclear matter),
as well as a rigid quark Fermi surface in a dense confining
matter, one indeed obtains a chiral restoration phase transition
within the confining matter. However,
if such a chirally symmetric confining 
phase exists at low temperatures, relevant degrees of freedom
near the Fermi surface are baryons. Quarks interact inside baryons.
Consequently, the quark distribution function near the "Fermi surface"
must be smooth. In this talk we report our findings for such
a diffused quark "Fermi surface" \cite{GSW}. Our main conclusion
is that for any reasonable diffusion there always exists such
a critical "Fermi momentum" of quarks at which the chiral restoration
phase transition persists.

\section{The model in a vacuum}

We use the $SU(2)_L \times SU(2)_R  \times U(1)_A \times U(1)_V $ 
symmetric hamiltonian with the instantaneous linear Coulomb-like
inter-quark potential.
This model was intensively used in the past to study chiral symmetry
breaking, chiral properties of hadrons, etc, \cite{Y}.
The model can be considered as a straightforward 3+1 dim
generalization of
the 1+1 dim 't Hooft model \cite{H}.
An important aspect of this model is that at large spins $J$
it exhibits the effective restoration of chiral symmetry in hadrons,
i.e., their mass is practically unrelated with the 
spontaneous breaking of chiral symmetry in the vacuum \cite{G2,WG}.

The self-energy of quarks,

\begin{equation}
\Sigma(\vec p) =A_p+(\vec{\gamma}\hat{\vec{p}})(B_p-p),
\label{SE} 
\end{equation}

\noindent
consists of the Lorentz-scalar chiral symmetry breaking part
$A_p$ and the chirally symmetric part $(\vec{\gamma}\hat{\vec{p}})(B_p-p)$.
The unknown functions $A_p$ and $B_p$ are obtained from the
gap equation. The functions $A_p$ and $B_p$, as well as the
self-energy of a quark,  contain an infrared divergence. Consequenty,
the single quark energy is infinite. At the same time this
infrared divergence cancels exactly in all possible color-singlet
quantities, like the quark condensate, hadron mass, etc, which is a
manifestation of confinement. 
While the model hamiltonian is manifestly chirally symmetric
(it does not contain any mass term), the self-interaction
of quarks produces chiral symmetry breaking via the non-zero
Lorentz-scalar self-energy $A_p$. This is how both confinement and
chiral symmetry breaking are guaranteed.

\section{Effect of a dense medium at $T=0$}

It is practically impossible to solve exactly the model in a dense
matter. Indeed, that would imply to solve it first for a single baryon; then
to obtain a baryon-baryon interaction; given this interaction to construct
a nuclear matter and then slowly to increase its density. Obviously,
it is a formidable problem. In order to proceed and get some
insight one needs justifiable simplifications.

In the large $N_c$ limit the nucleon is infinitely heavy, translational
invariance is broken and a many-nucleon system is certainly in a crystal
phase. Whether a (dense) nuclear matter will be a liquid or a crystal
at $N_c=3$ is a subject to dynamical calculations. Such
microscopical calculations cannot be performed for any "realistic" model in 3+1
dimensions with
confinement and (broken) chiral symmetry. However, in the real world 
$N_c=3$ we do know that the nuclear
matter is in a liquid phase; both translational and rotational
invariances are intact. We then assume a liquid phase with manifest
translational and rotational invariances in a dense quarkyonic matter.

 We treat the system in the mean field
approximation and assume first a simple valence quark distribution function,
like for the noninteracting quarks,
see Fig. 1.
Given this quark distribution function we solve the gap equation
and at some critical Fermi momentum obtain the chiral restoration
phase transition, see Fig. 2.

\begin{figure}[h]
  \begin{center}
    
      \center{\includegraphics[width=0.3\linewidth]{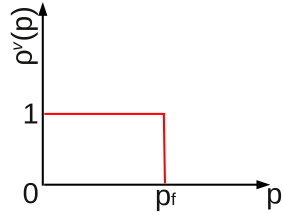} \\ }
    \end{center}
  \caption{Valence quark distribution for a rigid quark Fermi surface.}
\end{figure}

In a dense matter at $T=0$ the most important physics that leads
to  restoration of chiral symmetry is the Pauli blocking by
valence quarks of the positive energy levels required for the very
existence of the quark condensate. This is similar to the chiral
restoration in the Nambu and Jona-Lasinio model.
At sufficiently large Fermi momentum the gap equation does not
admit a nontrivial solution with broken chiral symmetry. Consequently,
the chiral symmetry breaking Lorentz-scalar part $A_p$ of the quark
self-energy vanishes and the chiral symmetry gets restored. 
However,
the chirally symmetric part of the quark self-energy does not vanish
and is still infrared divergent, like in the vacuum. This means that even with 
restored chiral
symmetry the single quark energy is infinite and the quark is confined.
This infrared divergence cancels exactly in all color singlet hadronic
modes that remain finite and well defined.

\begin{figure}
\begin{center}
 \includegraphics[width=0.5\hsize,clip=]{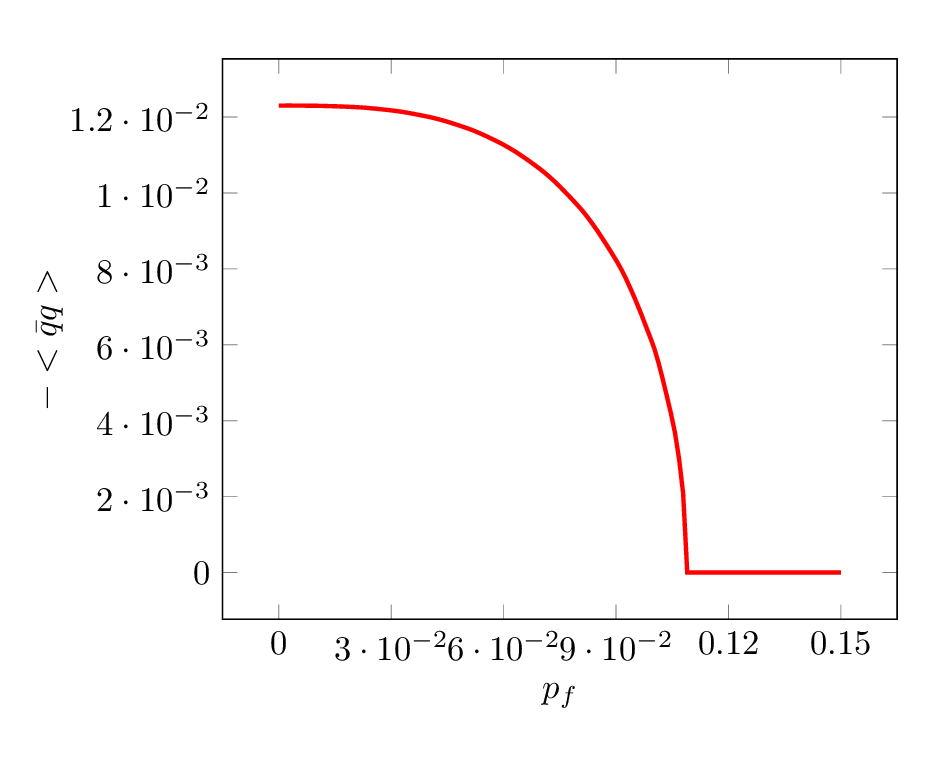}
 \caption{Quark condensate in units of $\sigma^{3/2}$
 as a function of the Fermi momentum, which is units of $\sqrt \sigma$.}
\end{center}
 \end{figure}

In this respect the model is radically different from the non-confining
NJL model. In the latter a dense matter is a Fermi gas of free
quarks. In the Nambu - Goldstone mode these quarks are massive.
In the Wigner-Weyl mode they are massless. In our case
physical degrees of freedom, that can be excited, 
are color singlet hadrons. In the
Wigner-Weyl mode these are the chirally symmetric hadrons.

\section{Diffusion of the quark Fermi surface}

In reality valence quarks near the Fermi surface interact and cluster into 
the color 
singlet baryons. This interaction in general would lead
to a diffusion of the rigid
Fermi surface for quarks. Some levels above the "Fermi momentum" must
be occupied with some probability as well as some levels below the
"Fermi momentum" with some probability must be empty. 

\begin{figure}[h]
  \begin{center}
    \includegraphics[width=0.3\linewidth]{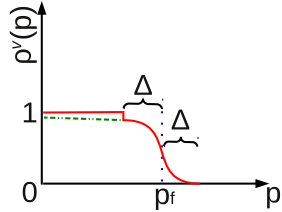} 
    \end{center}
  \caption{Valence quark distribution for a diffused quark Fermi surface.}
\end{figure}

In principle, the quark distribution function near the diffused Fermi
surface could be obtained self-consistently from the full solution of 
the problem.
It is a formidable task and such a program cannot be performed.
With the present state of the arts it is difficult
to obtain a microscopic insight into the dynamics
of the diffusion.
However, it is  clear that the realistic distribution function
will be smooth, of the form on Fig. 3.

We parameterize a smooth valence quark distribution function by
\begin{equation}
  \rho^v(p) = \Theta(-p + p_f - \Delta) +
  \Theta(p - p_f + \Delta)\frac{1}{e^{(p - p_{f}) / \Delta} + 1}. 
  \label{rho}
\end{equation}

\noindent
and solve the gap equation for
different $p_f$ and diffusion width $\Delta$.

For each fixed diffusion width $\Delta$ there always exists such
 critical "Fermi momentum" at which
 the chiral restoration phase transition does take place.
 This can be seen from Fig. 4, where a line of "critical
 Fermi momenta" is depicted. The area above this critical line
 corresponds to the chirally symmetric phase, while all points
 below the critical line represent a matter with broken chiral
 symmetry. 

\begin{figure}
\begin{center}
\includegraphics[width=0.5\hsize,clip=]{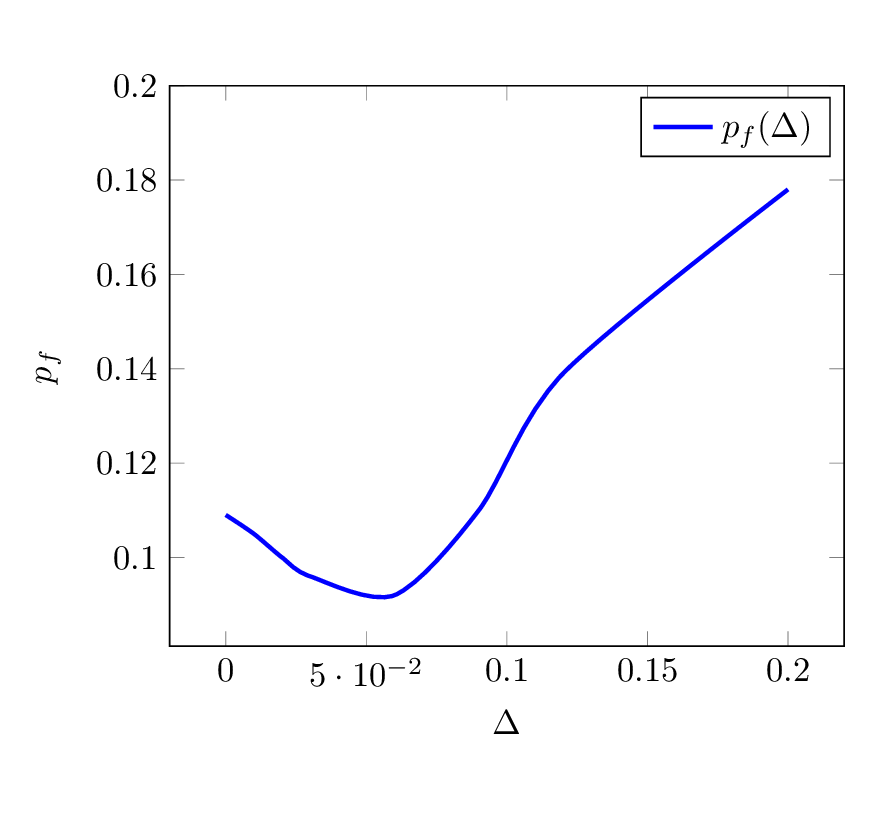}
\end{center}
\caption{Critical line that separates the quarkyonic matter
with broken and restored chiral symmetry.}
\end{figure}

This
can be easily understood. At all  momenta  $p \ll p_f$ the Pauli
blocking on Fig. 3 is the same as for the rigid quark Fermi surface.
At momenta just below the $p_f$ the effect of the Pauli blocking
is weaker than for the rigid Fermi surface. However, this is 
 compensated by additional Pauli blocking of the levels that
are just above the Fermi momentum for the rigid quark distribution.

This does imply that in a sufficiently dense strongly interacting
matter in the confining mode, assuming that it is a liquid phase,
the chiral restoration does take place. The mass generation mechanism
in such chirally symmetric but confining liquid is not related
to the chiral symmetry breaking.

Another explicit illustration of chiral symmetry 
of a dense matter  above the chiral restoration
phase transition are  properties of hadronic
excitations. In the Nambu-Goldstone mode of chiral symmetry
there must be a massless
excitation mode that is associated with the massless pion. At the
same time energies of all other  mesons must be finite.
In particular, there must be a finite splitting of the excitations
with quantum numbers $I, J^{PC} = 1, 0^{-+}$  and $I, J^{PC} = 0, 0^{++}$,
that will be referred as the pion and the $\sigma$-meson, respectively, according
to the standard nomenclature. In contrast, these excitations must be
exactly degenerate in the Wigner-Weyl mode of chiral symmetry and
form the $(1/2,1/2)_a$ representation of the $SU(2)_L \times SU(2)_R$
chiral group.

To obtain the quark-antiquark bound states we solve the homogeneous
Bethe-Salpeter equation in the rest frame.
\begin{figure}
\begin{center}
\includegraphics[width=0.8\hsize,clip=]{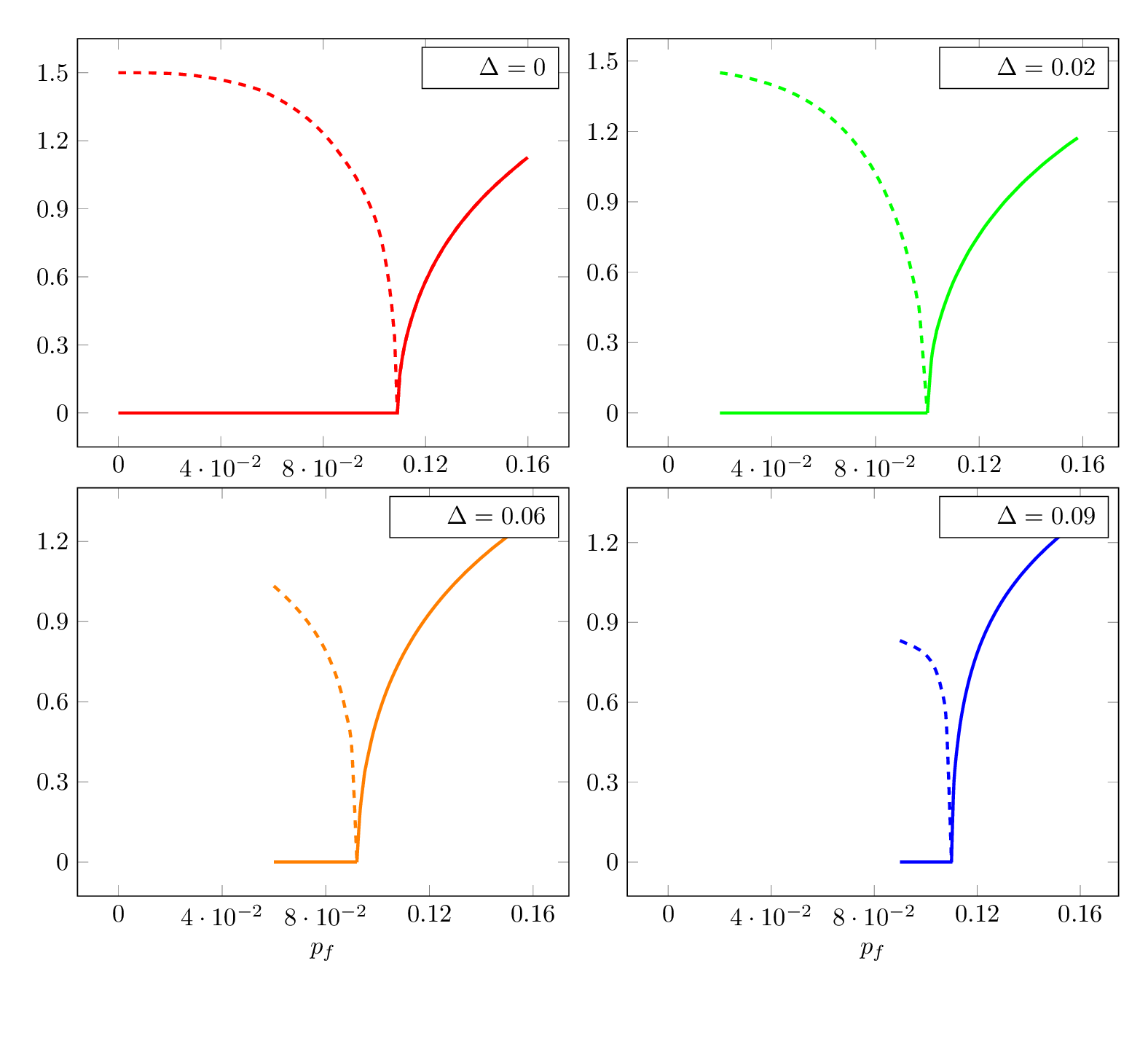}
\caption{Masses of the pseudoscalar (solid) and scalar (dashed) mesons 
in units of $\sqrt \sigma$ as functions of the "Fermi momentum" $p_f$
and of the diffusion width $\Delta$ (in units of $\sqrt{\sigma}$).}
\end{center}
\end{figure}
In the Wigner-Weyl mode, {\it i.e.}, when 
dynamical quark mass and chiral angle vanish, $M(p) = 0; \varphi_p = 0$, 
the Bethe-Salpeter equations
for the $1, 0^{-+}$ and $0, 0^{++}$ bound states
become identical  and consequently energies of these
states coincide.

On Fig. 5 we show masses of both pseudo-scalar and scalar modes
for different "Fermi momenta" $p_f$ and diffusion widths $\Delta$.
For each $\Delta$ there is a critical $p_f^{cr}(\Delta)$ at which the
chiral restoration phase transition takes place. Below this 
 $p_f^{cr}(\Delta)$ there is a massless pion and a massive $\sigma$-meson. 
 Above
the critical $p_f$ both the pion and the $\sigma$-meson are massive
and exactly degenerate.

\section{Conclusions}

In the confining mode the valence quarks interact and near the Fermi surface 
cluster into the color singlet baryons. This implies that there cannot
be a rigid quark Fermi surface and the valence quark distribution
function near the Fermi surface must be smooth. 
 Assuming a liquid phase, i.e., unbroken translational and
rotational invariances, we
parameterize such  diffused "Fermi surface" by a
simplest possible function and solve the corresponding gap and
Bethe-Salpeter equations. By this we  verify whether a chiral phase transition,
previously observed for a rigid quark Fermi surface, survives or not.
It turns out that for any reasonable diffusion width there always exists
such a "Fermi momentum" that the chiral restoration phase transition
does take place. This reconfirms our previous conclusions about
possible existence of the confining but chirally symmetric phase.
Below the phase transition the elementary excitation modes of a matter
are hadrons with broken chiral symmetry, while above the phase transition
such excitations are chirally symmetric hadrons.

\bigskip
{\bf Acknowledgements}
Support of the Austrian Science
Fund (FWF) through the grant P21970-N16 is acknowledged.

\end{document}